# Infrared Spectroscopy of HR 4796A's Bright Outer Cometary Ring + Tenuous Inner Hot Dust Cloud

C.M. Lisse[1], M.L. Sitko[2], M. Marengo[3], R. J. Vervack Jr.[1],
Y.R. Fernandez[4], T. Mittal[5], C.H. Chen[6]

[1] JHU-APL, 11100 Johns Hopkins Road, Laurel, MD 20723  carey.lisse@jhuapl.edu, ron.vervack@jhuapl.edu

[2] Department of Physics, University of Cincinnati, Cincinnati, OH 45221-0011 and Space Science Institute, Boulder, CO 80301, USA  sitkoml@ucmail.uc.edu

[3] Department of Physics and Astronomy, 12 Physics Hall, Iowa State University, Ames, IA 50010  mmarengo@iastate.edu

[4] Department of Physics, University of Central Florida, Orlando, FL 32186-2385  yan@physics.ucf.edu

[5] Department of Earth and Planetary Sciences, McCone Hall, University of California at Berkeley, Berkeley, CA 94720  tmittal2@berkeley.edu

[6] Space Telescope Science Institute, 3700 San Martin Dr. Baltimore, MD 21218  cchen@stsci.edu.

26 Pages, 5 Figures

# Abstract

We have obtained new NASA/IRTF SpeX spectra of the HR 4796A debris ring system. We find a unique red excess flux that extends out to ~9 μm in *Spitzer* IRS spectra, where thermal emission from cold, ~100K dust from the system's ring at ~75 AU takes over. Matching imaging ring photometry, we find the excess consists of NIR reflectance from the ring which is as red as that of old, processed comet nuclei, plus a tenuous thermal emission component from close-in, T ~ 850 K circumstellar material evincing an organic/silicate emission feature complex at 7 - 13 um. Unusual, emission-like features due to atomic Si, S, Ca, and Sr were found at 0.96 - 1.07 μm, likely sourced by rocky dust evaporating in the 850 K component. An empirical cometary dust phase function can reproduce the scattered light excess and 1:5 balance of scattered vs. thermal energy for the ring with optical depth $\langle\tau\rangle \geq 0.10$ in an 8 AU wide belt of 4 AU vertical height and $M_{dust} > 0.1\text{-}0.7\ M_{Mars}$. Our results are consistent with HR 4796A consisting of a narrow sheparded ring of devolatilized cometary material associated with multiple rocky planetesimal subcores, and a small steady stream of dust inflowing from this belt to a rock sublimation zone at ~1 AU from the primary. These subcores were built from comets that have been actively emitting large, reddish dust for > 0.4 Myr at ~100K, the temperature at which cometary activity onset is seen in our Solar System.

# 1. Introduction

HR 4796A is a star of spectral type A0.5V (effective temperature $T_{eff}$ ~ 9350 K, $M_*$ = 2.2 $M_{Sun}$, $R_*$ = 1.7 $R_{Sun}$, $L_*$ = 23 $L_{Sun}$, [Fe/H] = -0.03) located at 72.8 ± 1.7 pc, with estimated age of 9 ± 2 Myr and a known infrared (IR) excess (Jura 1991, Stauffer *et al.* 1995, Saffe *et al.* 2008; Chen *et al.* 2006, Cameron *et al.* 2015, Donaldson *et al.* 2016). Together with the 3-magnitudes fainter M2.5V star HR 4796B, it forms a binary system with a projected separation of ~560 AU. Its young age places HR 4796A at a transitional stage between massive gaseous protostellar disks around young pre-MS T Tauri and Herbig Ae/Be stars (~1 Myr) and evolved and tenuous debris disks around MS "Vega type" stars (~100 Myr; see Jura *et al.* 1993; Chen & Kamp 2004). With a fractional IR luminosity of 5 × 10$^{-3}$, the disk around HR 4796A is one of the brightest cold debris disk systems among main-sequence stars. The dust in the disk has been resolved in the mm with ALMA and SCUBA (Greaves *et al.* 2016), at mid-infrared (MIR) wavelengths (Koerner *et al.* 1998; Jayawardhana *et al.* 1998), at near-infrared (NIR) wavelengths using HST (Schneider *et al.* 1999, Lagrange *et al.* 2012) and from the ground, with adaptive optics and coronography (Mouillet *et al.* 1997; Augereau *et al.* 1999; Thalmann *et al.* 2011; Lagrange *et al.* 2012; Wahhaj *et al.* 2014; and Perrin *et al.* 2015). The dust was also resolved at visible wavelengths with HST by Schneider *et al.* (2009), who reported a disk confined to a ring with maximum brightness at its inner edge located at ~75 AU from the primary, a radial extent of less than 17 AU, a sky orientation with the ring long axis at a position angle (PA) of 26.8° and an inclination of 75.8° with respect to pole-on. The east side was observed to be brighter than the west side, which led to the conclusion that the east side was inclined towards us under the common assumption of preferentially forward-scattering grains (although this is disputed by recent GPI coronographic studies (Perrin *et al.* 2015). HST/NICMOS high-angular-resolution images of this system show that the disk ring is narrow, with steep inner and outer edges (Lagrange *et al.* 2012).

HR 4796A has been studied in detail for over 20 years, but there is still debate about what constitutes the circumstellar material surrounding it. Early modeling (Augereau *et al.* 1999) showed that two components are needed to explain the spectral energy distribution (SED) up to 850 μm and the resolved images up to mid-IR: a cold component, corresponding to the dust ring observed in scattered light, made of icy and porous amorphous silicate grains in sizes up to a few meters in radius massing > 2-3 $M_{Earth}$ *in toto* and dust-to-gas mass ratio (D/G) > 1, plus a hotter component closer to the star, with properties similar to micron-sized cometary refractory silicates. The need for the hotter component was debated by Li & Lunine (2003), but mid-IR images between 8 and 25 μm with improved spatial resolution argued for the presence of hot dust within ~10 AU from the star (Wahhaj *et al.* 2005). Debes *et al.* (2008) disagreed, proposing reflectance from a massive population of dust grains dominated by 1.4 μm tholins (a complex organic material, detected as a major constituent of the atmospheres and surfaces of Titan and Pluto) to explain their 7-band, 0.6 to 2.2 μm spectrophotometry (which is characterized by a steep red slope increasing from λ ≈ 0.5 to 1.6 μm followed by a flattening of the spectrum at λ > 1.6 *μm*). Kohler *et al.* (2008) showed that reflectance from porous grains composed of common cosmic dust species (amorphous silicate, amorphous carbon, and water ice) could reproduce those data as well. Hinkley *et al.* (2009) used imaging polarimetry and simple morphological models with an empirical scattering phase function and Rayleigh-like polarisability to conclude that their measurements are compatible with a population of micron-sized dust. More recent measurements of the far-infrared excess emission with APEX (Nilsson *et al.* 2010) and Herschel (Riviere-Marichalar *et al.* 2013) confirmed the presence of a cold dust component and were able to constrain its mass. The mass in dust grains below 1 mm was estimated at 0.15 $M_\oplus$, using the flux density at 870 μm by Nilsson *et al.* (2010). The most recent ALMA measurements confirm the presence of a massive mm particle outer dust ring, but find no associated CO gas (Greaves *et al.* 2016).

Thalmann *et al.* (2011) used Subaru/HiCIAO high-contrast imaging of the cold dust ring to discover a significant offset between its center and the star's position, similar to the situation in the older (~300 Myr) A3V Fomalhaut system (Kalas *et al.* 2013; although no planet candidates were detected near HR 4796A, and the upper limits to their size were < 10 $M_{Jup}$). Planets are now expected in the system, as they can generate the observed disk asymmetries through gravitational confinement or perturbation (Wyatt *et al.* 1999) and form rings with sharp edges (Augereau *et al.* 1999; Klahr & Lin 2001; Thebault & Wu 2008). In 2015, Perrin *et al.* used Gemini Observatory's GPI to acquire coronographic H-band imaging, confirming the unusual sharpness and narrowness of the ~75 AU ring, but they found difficulties reconciling SED-based modeling with a theoretical scattering phase function. Milli *et al.* (2017) produced the run of photometry for the entire ring at from 0.96 to 1.67 μm, finding a pronounced brightness asymmetry inconsistent with small spherical Mie particles but consistent with a disk composed of large porous aggregates dust with radius > 1 μm, and no unseen planets above 1.5 to 15 $M_{jup}$, (depending on separation from the primary).

In this work we present a new 0.8 – 5.0 μm, R ~ 1000 spectrum of the HR 4796A system taken as part of the 100+ hours Near InfraRed Debris disk Survey (NIRDS; Lisse *et al.* 2017; 30 systems to date) of northern debris disks using the NASA/IRTF 3m's SPEX instrument (Rayner *et al.* 2003, 2009; Vacca *et al.* 2003, 2004; Cushing *et al.* 2004) to help resolve the debate concerning the nature of the circumstellar material in this system. Our spectrum is novel in that it covers the entire wavelength range from 0.8 - 5.0 μm; previous studies have only used small snippets of this range and have not gone past 2.5 μm. The HR 4796A spectrum with its strong linear red excess from 2 to 5 μm is also unique amongst the NIRDS dataset, including the 15 star systems that evince SEDs produced by the stellar photosphere + cold outer system dust and 3 YSO/Transition disks of similar age to HR 4796A. In conjunction with Spitzer 5.3 - 35 μm IRS R ~ 100 spectroscopy (Chen *et al.* 2006, 2014; Mittal *et al.* 2015) reduced using the latest calibration software, we present a combined SED that demonstrates an excess above the primary's photospheric emission due to a combination of scattering at 1- 9 μm and thermal emission from 10 - 35 μm from the cold dust ring at 75 AU plus thermal emission from a tenuous, unimaged hot inner component at ~850K and ~1 AU.

## 2.    Observations

We observed the HR 4796A system at 2.2 - 5.0 μm on 04 Feb 2012 at 13:56 UT, and from 0.8 – 2.5 μm on 30 May 2013 at 05:44 UT, from the NASA/IRTF 3m telescope on the summit of Mauna Kea, HI.  Prior to an upgrade in 2014, the SpeX instrument provided R = 2000 to 2500 observations from 0.8 – 5.0 μm in two orders, termed SXD (for "short cross-dispersed") and LXD2.1 (for "long cross-dispersed"), when configured to use an 0.3"slit and the standard optics chain (Rayner *et al.* 2003; 2009). The seeing was ~1.2" for both sets of HR 4796A observations, large enough that light from the star + disk were convolved together in the slit. Our observational setup was identical to that used for other NIRDS debris disk studies (e.g. Lisse *et al.* 2012, 2015, 2017). The nearby A0V star used in our Spextool data reduction as a calibration standard (Vacca *et al.* 2004) was the *K* = 6.87 star HD 112680, picked to match the *K* = 5.77 HR 4796A system in brightness, color, effective temperature and spatial proximity (the best-fit BtVtJHKs photometry gives Teff = 9450K and (B-V)$_o$ = -0.022, consistent with an unreddened A0V star (Pickles & Depagne 2010) that is Δ(B-V) ~ 0.02 redder than HR 4796A). We observed both stars in SXD mode with a total on-target integration time of 300 sec, and in LXD mode for a total time of 300 sec. The instrument behavior was nominal throughout each set of observations and the summit weather was good, allowing for good sky correction and stellar calibration. Both stars were observed in ABBA nod mode to remove telescope and sky backgrounds. We observed HR 4796A before morning twilight and after meridian transit in 2012, with the position angle (P.A.) of the slit on the sky set at ~ 357$^o$ in order to be perpendicular to the horizon; and in 2013, after sunset near transit with the slit at P.A. ~345$^o$. The M2.5V-star companion HR 4796B (*K* = 9, separation 7.7" at P.A. 226$^o$) would not have been in the slit during our observations, and thus did not contribute significant flux to our measurements.

## 3.     Results

Figure 1a shows the entire grasp of our HR4796A SpeX measurements in the context of 6 other NIRDS main sequence early A-type stellar spectra. The typical spectral behavior for our program systems matches the photospheric model well from 0.8 - 1.3 um, exceeds it slightly (if at all) from 1.3 - 3 um, then matches it again from 3 - 5 um. For a small subset of ~8 young NIRDS stars, with abundant warm (200 - 500 K) circumstellar dust, we find a close match to a stellar photospheric spectrum out to ~3 um, then a rapid, exponential increase in excess from 3 to 5 um. HD 131488 shows the typical growing exponential behavior seen for these NIRDS stars (e.g., HD 113766, HD 15407A, HD 23514; Lisse *et al.* 2012, 2015) dominated by warm (~300K) thermal dust emission observed on the Wien law side of its SED. The spectral behavior of HR 4796A in the NIR looks like neither of these behaviors, but instead evinces its own unique slowly, almost linearly, growing excess from 2 to 5 um.

Figure 1b shows a small region of the HR4796A SpeX measurements, from 0.96 - 1.07 μm, in detail compared to the other A-stars of Figure 1(a). This region was chosen because unusual *emission-like* features due to Sr II and S I appear in the HR4796A spectrum where only absorptions are seen for the other A-stars. Further, there is some evidence (albeit weak and near the SNR limit of the data) that the normal absorption features for Ca II and Si I are filled in by additional emission in HR4796A. At longer wavelengths, however, we find that there are no detectable HBrγ and CO emission lines due to fluorescing circumstellar gas or accretion, or HeI 1.083 μm, FeII 1.256/1.644 μm, and $H_2$ S(1) 1-0 2.121 μm emission lines due to strong wind outflow (Connelley & Greene 2014) within the noise of the measurement, so we can rule out the presence of a large scale circumstellar gas disk surrounding the primary.

Our 0.8 - 5.0 μm SpeX results for HR 4796A are shown in detail in Figure 2a in blue, combined with a re-extracted and calibrated Spitzer IRS spectrum of the system originally taken by Chen *et al.* (2006, 2014) but re-extracted for this work using the latest fringe-removal software. The independently calibrated spectra agree well, and are consistent with ALLWISE, AKARI, and Spitzer photometry of HR 4796A, and with saturation-corrected 2MASS and Tycho synthetic photometry (*Vizier* photometric database http://vizier.u-strasbg.fr/vizier/sed/; Pickles & Depagne 2010). The overall appearance is of a combined SED dominated by stellar photospheric emission at the shorter wavelengths, with a small but growing excess flux extending out to ~9 μm, then a rapid increase in flux due to a cold thermal dust source peaking around 30 *μm*. Subtracting off our best model fit to the A0.5V's photosphere (yellow in Figure 2a), we find the residuals shown as black points at the bottom of the plot. The photospheric model absolute normalization was chosen so that our residuals match the 0.8 - 2.2 μm Rodigas *et al.* 2015 and Milli *et al.* 2017 total ring photometry (black circles; this is discussed more in Section 4). Our residuals across the 0.8 - 2.2 μm datasets are roughly consistent with the spectrum of a 7500 K blackbody (red curve in Fig. 2a). There is, however, a marked jump in our residuals versus the 3 - 3.4 μm ring photometry published by Rodigas *et al.* (2015). The source of this jump becomes clear when we also subtract a 100K blackbody model from the combined SED, and find what appears to be the signature of an ~850 K blackbody + silicate emission feature at 8 - 13 um.

In Fig 2b we show an extended view of HR 4796A's SED (blue curves for the IRTF & Spitzer data, triangles for literature photometry) in $\nu F_\nu$ space, with the short and long wavelength excesses fit by 7500 K and 100K blackbodies, respectively. The ratios of the areas under these blackbody curves indicate an NIR excess of ~0.2% and an FIR excess of ~0.8% of the primary star's luminosity.

The NIR excess could be due to a body with color temperature of 7500 K; but we could also interpret the excess as due to scattering of the primary's starlight, i.e., $F_{observed}(\lambda) = F_{photosphere}(\lambda) +$ Reflectance$(\lambda)*F_{photosphere}(\lambda)$, in which case we find the interesting and very red reflectance trend shown in Figure 3. This is interesting because searching through the known literature for reflectance measurements of outer bodies in our solar system, the only match we find for this reflectance is from the surface of old, highly processed and devolatilized comet nuclei (Figure 4).

Which interpretation of the NIR excess is correct? The literature on HR 4796A contains conflicting model results supporting both, for quite plausible reasons owing to the limited data used to constrain previous models. Using our new SpeX spectrum covering the entire wavelength range of 0.8 - 5.0 μm, we can help clarify the situation. We can rule out our NIR excess at 0.8 - 2.2 μm as due to hot dust close to the star (Koerner *et al.* 1998, Augereau *et al.* 1999) unless this dust is somehow stable at T~ 7500 K, contrary to our understanding of the condensation of solid materials in solar systems at < 1800 K (see, for example, Yoneda & Grossman 1995, Ebel 2006, Davis & Richter 2014 and references therein), and was somehow undetectable by HST/NICMOS (Schneider *et al.* 1999, Lagrange *et al.* 2012) and GPI (Perrin *et al.* 2015). Similarly, we can rule out the NIR excess sourced by a close-in small stellar or sub-stellar companion, because a point source object with ~ 0.2% of the primary's luminosity and of its 2 μm flux would have been detected by GPI in the NIR (Perrin *et al.* 2015). Similarly, a close-in low mass young M-star or brown dwarf companion would have been easily detected in the x-ray, as HR 4796B was by *Chandra* (Drake *et al.* 2014). Finally, Hipparcos long-term astrometry (van Leeuwen 2007) shows no evidence for perturbations due to an unresolved companion.

**Inner Hot Dust.** A NIR ring scattering excess + FIR thermal ring excess model held together beautifully until we noticed that there was an apparently large discrepancy between our calculated 3.0 - 3.5 μm flux residuals and those reported by Rodigas *et al.* (2015) using LBT/CLIO measurements. Our first reaction was to dismiss the discrepancy due to the large reported uncertainties on the CLIO broad-band measurements compared to our detailed spectral observations. But then we noted that there was "room", after allowing for the outer ring's reported short wavelength flux as scattered light, for a high temperature (~850 K) inner system thermal component in the SED at the $f_{excess}/f_* \sim 10^{-4}$ level that would add flux to our total measurement but not to any discrete ring measurements (Fig 2b). We went looking for confirmation of this in the Spitzer data by removing our best blackbody model for the ~100K cold dust thermal emission (Fig 2), and found a spectral signature at 7 - 13 μm consistent with thermal emission from a mix silicates and refractory carbonaceous species (Figure 5). We did NOT find the expected strong associated silicate emission features at 17 - 23 μm, which we should have easily seen if the silicates were in the cold dust, verifying that 7 - 13 μm features were from the hot inner component. Looking in detail at the A-star spectral comparison of Figure 1, we find that HR4796A evinces small but detectable emission-like features where the other A-stars demonstrate absorption features for lines of S I, Ca II, Sr II, and Si I (but not FeI, MgI, or CI). Emission line features in disk systems are usually associated with the presence of fluorescing circumstellar gas species, and 850 K is in the range of the evaporation temperatures for moderately refractory metal sulfides & Ca/Sr-rich pyroxene silicates, and thus could produce the atomic emissions seen. Scaling using a $T^4$ blackbody total luminosity law, we estimate that the surface area of inner hot dust required to produce the putative 850 K flux is very small, ~7 x $10^{-6}$ that of the outer ring, which could easily be lost in the noise of any ring + star imagery.

In summary, this tenuous inner hot component, similar to the hot dust excesses reported by Augereau *et al.* (1999) and Wahhaj *et al.* (2005) (or, if unresolved, the unusual stellar flux by Jura *et al.* 1995), would require ~$10^{-5}$ the surface area of the ~100 K cold dust component to create the estimated amount of flux, could produce an observable 10 μm silicate line complex and a 20 μm complex that is too small to be seen beneath the overwhelming 20 μm emission from the cold ring dust, and at ~850K is in the range of the evaporation temperatures for metal sulfides and Ca/Sr-rich pyroxenes and thus could produce a region of enhanced dust surface area via sublimation fragmentation and the S I, Ca II, Sr II, and Si I atomic emissions seen. Ferromagnesian sulfide and pyroxene-rich dust is found in primitive cometary bodies in our solar system (Lisse *et al.* 2007a,b; 2008) and in the primitive dust of the η Corvi system (Lisse *et al.* 2012, Fig. 5), so it is quite plausible that the outer dust ring is sourcing the inner hot dust cloud via P-R drag. At ~$10^{-8}$ the surface area of the cold dust ring, it could also be non-detectable vs. the star + ring in optical or short wavelength NIR imaging observations of the system. We require this new inner hot component to reconcile the difference between our star- and ring-thermal flux removed residuals and Rodigas *et al.* 2015's ring photometry (green curve in Fig 2).

The new ~850K MIR excess is at ~0.03% of the primary's luminosity, faint versus the primary star's emission, but clearly discernable in our high precision spectroscopy. It can explain previous determinations of hot dust emission near the star (e.g., Jura *et al.* 1993; Augereau *et al.* 1999; Wahhaj *et al.* 2005). Its faintness makes its detection problematic, resulting in the controversy over its existence in previous studies.

## 4. Discussion

**Novel NIR Spectral Data.** The power of characterizing circumstellar dust systems in the near-infrared using medium-resolution spectroscopy can be seen in Figure 1. Using more than 6000 independent spectral data points with ~1% relative precision, we can accurately apply photospheric models at the percent level without encountering the usual 5-10% calibration uncertainties entailed in photometric fitting using TYCHO, MICHIGAN, WISE, IRAC, and 2MASS catalogue data. Doing so, we can search for the emergence of excess emission as we go from the shortest wavelengths where the star's flux dominates out to 5 um, where thermal emissive effects from circumstellar material becomes important. (Similar results were found for NIRDS F-stars observed with SpeX, and published in Lisse *et al.* 2012 and 2015.) The NIRDS spectrum for HR4796A shows a clear flux excess at the few % level over photospheric emission that grows with increasing wavelength, although the growth is much much slower than the exponential Wien law rise we see in the warm (~200-400K) dust dominated systems we have observed (HD113766, HD145263, HD98800B, etc.; Lisse *et al.* 2015, 2018) The SpeX spectral slope also matches up well at 4 - 5 μm with the published Spitzer/IRS slope at 5.3 - 9.0 μm (Chen *et al.* 2014, Mittal *et al.* 2015), and to archival AKARI, WISE and Spitzer/IRAC photometry (Figure 2), showing that the NIR excess continues out to longer wavelengths before the thermal emission from the cold, ~100 K dust in the circumstellar ring takes over beyond 10 um.

**Photospheric Model Normalization.** The largest uncertainty in our analysis is the determination of the absolute level of the stellar photospheric flux. To this end we have performed a model fit to the continuum and absorption line depths in the 0.8-1.25 μm region using NEXTGEN stellar models (Allard *et al.* 2012, with solar abundances from Asplund *et al.* 2009). Assuming solar metallicity (reported values of [Fe/H] for HR 4796A are = -0.03; Saffe *et al.* 2008) we find a best-fit model with $T_{eff}$ = 9350 ± 200 K (2σ), log g = 4.05 ± 0.26 (2σ) (Fig 2). This corresponds to a solar-type A0.5V star, consistent with literature reports of HR4796A being either an A0V (e.g., Schneider *et al.* 2009) or an A1V (e.g. Mamajek 2016). An equivalent ATLAS9 model (Castelli *et al.* 1997), calculated with the same best fit $T_{eff}$ and log g parameters, produced indistinguishable results.

Normalizing the best-fit model equal to the observed SpeX flux at ~0.81 μm (Fig 2), we found a lower-limit infrared excess while satisfying the requirement that the model photospheric flux is less than or equal to the observed flux from 0.8-100 μm within the uncertainties of the observations. This produced a total excess flux of color temperature ~3100 K, but negligible excess flux from 0.8 - 1.2 μm, contradictory to literature detections of the ring in the optical and JH bands. We instead adopt a model photospheric normalization that is 0.1% smaller, which produces residual flux over the photosphere that matches the run with wavelength of the outer ring photometry surface brightness at 90° phase reported by Rodigas *et al.* (2015) (and verified in the total ring photometry measurements of Milli *et al.* (2017)), while still fitting the 2MASS photometry within 2σ limits. The result, with model excess color temperatures ~7500K, is shown in Fig. 2. Fortunately, the normalization for this model is only slightly (0.1%) lower than for the minimum excess case. By arguing that even the most conservative excess normalization case, at ~3100 K, has a color temperature much higher than ever seen for solid astrophysical dust, we are able to rule out the NIR excess from 0.8 - 2.2 μm as sourced by thermal dust emission.

**Outer Ring.** After allowing for the flux contribution from the inner hot component, we conclude that the bulk of HR4796A's NIR excess is due to scattering of ~0.2% of the primary's light from the ~75 AU outer dust ring. This is in association with an ~100 K MIR/FIR thermal emission excess from the same ring material, at an approximately equal amount of total re-emitted energy. The scattered ring light reflectance is extremely red (Fig. 3), and demonstrates some flattening at 1.5 - 1.9 μm followed by a re-steepening at longer wavelengths up to 9 μm. Broad absorptions at ~3.4 μm due to the cometary C-H organic stretch (Quirico *et al.* 2016) and another predicted by Debes *et al.* at 3.8 - 4.0 μm for tholin material are also possibly present (Fig. 3), but our excess reflectance spectra are noisy in this region due to telluric atmospheric effects and we believe they need future verification. The lack of any silicate emission above an ~100 K thermal continuum (Fig. 2a,b) argues that any rocky dust particles present must be optically thick, with average radius $a_{grain} \geq 10$ μm (if bare) or containing a thick overcoat of red organic material ($a_{coat} > 3\ \mu m$) (Lien 1990, Lisse *et al.* 1998). Both of these minimum size ranges are consistent with the a ≤ 5 μm radiative blowout instability for the system (Jura *et al.* 1993, 1995), implying that the outer ring is not brand new, nor is it actively creating significant amounts of micron-sized particles. At 70 to 100 μm (Fig. 2b), we see a distinct falloff of the Spitzer MIPS, IRAS, and AKARI excess thermal emission vs. the best-fit blackbody model (T = 100 K assuming 1 blackbody, with a broad poorly fit residual peaking at ~16 μm. A slightly better two-blackbody fit has T = 95 K as the dominant emission source, and T = 130 K as a secondary minor source with ~4% the optical depth of the 95 K source). This behavior is similar to the emissivity falloff seen for cometary dust at wavelengths where the size distribution is dominated by particles with radius < wavelength (Lisse *et al.* 1998); essentially the particles become worse and worse antennas as the wavelength increases. We can thus place an upper size bound for the dust dominating the ring population of a ≤ 50 μm.

But why then is the HR 4796A ring material producing such a strongly red scattering excess? There are 15 other photosphere + cold Kuiper Belt dust systems in the NIRDS survey that do not evince such behavior, including Fomalhaut (A1-3V) and HD 32297 (A4V) and three other early-to-mid type A stars of high luminosity (Fig. 1). Six other NIRDS warm dust systems with dominant 10 μm silicate complexes show excesses in the 3-5 μm region, including the A3V star HD131488 (Fig. 1), but these have the characteristic spectral shape of the Wien law tail for 300 - 400 K dust. A clue may be found by comparing our reflectance spectrum to that of the reddest Solar System objects: the small bodies, Centaurs, and KBOs found in our outer Solar System (Fig. 4). The red spectrum best matches that of ***active*** short period comets observed in our Solar System (Fig. 4), consistent with the modeling of Li & Lunine (2003). It definitely does not match KBO or distant, inactive Centaur spectra, and only matches that of the active and famously red Centaur 5145 Pholus out to ~2 μm - these objects have surface reflectance spectra dominated by spectrally flat and neutral ices at longer wavelengths (Barruci *et al.* 2003, Jewitt & Luu 2004).

This is at first surprising to find, but makes sense when we note that the temperature of the cold dust thermal emission in the SED is ~95 – 100 K, the temperature at which two cometary activity drivers in our Solar System become important (Meech *et al.* 2013; Kiss *et al.* 2015; Jewitt 2009): the onset of amorphous water ice crystallization and the crossing of the $CO_2$ "ice line". Bluish volatile ices, such as $N_2$, CO, $CH_4$, $NH_3$ and $CO_2$, are thus not stable on these bodies' surfaces, as they are on the more distant inactive Centaurs and KBOs. Instead these ices are actively removed, exposing the ultra-red (in the NIR) and thermally stable organic rich material beneath. It takes an A0V star with $L_* \sim 23\ L_{Sun}$, and slowly rotating bodies to reach these temperatures at 75 AU distance. The expected D/G in the disk should be that of comets, $\geq 1$, safely higher than the HST limits of D/G $\geq$ 0.25 found by Chen & Kamp (2004) or D/G $\geq$ 10 found by ALMA (Greaves *et al.* 2016). A silicate + organics mixture of materials, as favored by Rodigas *et al.* 2015's analysis of the scattering + thermal properties of the HR4796A excess, is exactly what one expects for devolatilized comet dust (Lisse *et al.* 2006, 2007, 2012). One other piece of evidence we have is the narrowness of the HR 4796A ring: this system appears to be dynamically well constrained to an annulus 8 AU in width (Perrin *et al.* 2015) and thus the ring formed sometime ago (where "sometime ago" is > 1000 orbits (Thebault & Wu 2008), or > 0.4 Myr for objects at 75 AU orbiting a 2.2 $M_{Sun}$ primary), which allows time for cometary surface reddening via the primary's strong UV flux and for the blowout of any small (and thus relatively blue) dust particles from the system via its high $L_*$ (Jura *et al.* 1993, 1995). This timescale fits nicely with estimates of a comet's lifetime vs. mass depletion at 100K in our Solar System, which are $M_{comet}/Q_{dust} = 10^{13}$ kg/0.3 kg/sec = 3 x $10^{13}$ sec ~ 1 Myr (Farnham *et al.* 2014, Farranochia *et al.* 2014, Tricario *et al.* 2014).

**HR4796A System Models.** The total and relative amounts of energy emitted by HR 4796A also put important constraints on any successful model of the system. In order to produce the absolute amount of scattered energy of ~0.2% of the star's total luminosity plus a thermally re-emitted excess flux of ~0.8%, the cometary material in a ± 4 AU high belt (i.e., in a belt with assumed vertical scale height of 4 AU) at 75 AU must have an average optical thickness $<\tau> \geq 0.10$ so as to geometrically intercept enough starlight, which is very high compared to the typical $<\tau> = 10^{-3}$ to $10^{-6}$ values seen in cometary comae. (We thus note in passing that this scenario is similar to the optically thick "3rd model" proposed by Perrin *et al.* (2015) to explain their GPI scattered light and polarization measurements in the 1.6 - 2.2 μm region.) The ratio of scattered energy/(scattered + thermal emission energy) is approximately 0.25 (Fig 2b), which is unusual for low geometric albedo cometary dust. (E.g., $f_{Scatt}/(f_{Scatt}+f_{Therm})$ = 0.065 for comet C/Austin 1989 dust at 92° phase and = 0.13 for C/Levy 1990 dust at 69° phase (Lisse *et al.* 1998)). However, there is a very strong scattering enhancement due to our nearly edge-on viewing geometry of HR 4796A, at the very low scattering angles that cometary dust is not usually observed at in the infrared. Fig 2b (inset) shows that we can expect an enhancement of a factor of as much as 8 over the face-on case for an $i = 17^o$ ring composed of cometary dust with a phase function similar to that found by Schleicher (2010), easily enough to account for the large proportion of scattered light energy.

In summary, a plausible case can be made that HR 4796A is special in that within the first ~9 Myr of its existence it has created an extraordinarily narrow and dense ring of reddish devolatilzed cometary dust at ~75 AU from the HR 4796A primary at a local effective temperature of ~ 100K. The cometary nuclei sources have been actively outgassing, driven by either water ice crystallization or $CO_2$ sublimation, for ~ 1 Myr. What is left over is similar to dust emitted from comets in our Solar System, i.e., darkish, red, and made up of devolatilized, large, porous, fluffy dust aggregates composed of amorphous carbon, ferromagnesian silicates, and ferromagnesian sulfides (Lisse *et al.* 2006, 2007; Sitko *et al.* 2011, Egrand *et al.* 2016). This composition is consistent with the findings of Li & Lunine (2003), Koheler *et al.* (2008), Nilsson *et al.* (2010), Rodigas *et al.* (2015), and Milli *et al.* (2017). The dust includes a range of particle sizes with $dn/da \sim a^{-3}$ to $a^{-4}$ (Lisse *et al.* 1998, 2004, 2005; Green *et al.* 2008; Kelley *et al.* 2014) dominated by 5 - 50 μm radius particles. The effects of radiation pressure should efficiently sort and remove any small, a < 5 μm particles, while leaving the larger particles intact to slowly spiral in via the influence of Poynting-Robertson drag. The ring must be warm enough to have produced cometary activity, devolatilization, and processed dust emission, but not so warm that it depleted the cometary population in less than 0.4 Myr of insolation-driven activity. The average optical depth along a radial line of sight, $<\tau>$, through the ring must be greater than 0.10 in the ± 4 AU region above and below its midplane in order to intercept enough starlight and produce the scattered light and thermally re-emitted fluxes of Figure 2. This in turn implies a minimum ring mass of $0.10 * 2\pi rh * a_{min} * \rho_{dust} = 0.2 * \pi * 75$ AU $* 8$ AU $* 5$ μm $* 2.5$ g/cm$^3$ = 1.0 x $10^{23}$ kg or 0.16 $M_{Mars}$, assuming the lowest possible dust particle size of 5 μm with density of 2.5 g/cm$^3$ due to a mixture of rocky silicates and coated with organics. With a more realistic $dn/da \sim a^{-3}$ comet particle size distribution for particles 5 - 50 μm in size, the mass required for the same extinction cross section increases by a factor of 10/ln(10) = 4.3, and the minimum ring mass is 4.5 x $10^{23}$ kg or 0.7 $M_{Mars}$.

The observed cold dust ring structure is extremely narrow and dynamically "cold", a difficult thing to understand in a young developing solar system. The velocity dispersion in a young planetesimal belt should be high compared to the $V_{Keplerian}$ = 5 km/sec velocity of ring particles at 75 AU from the 2.2 $M_{Sun}$ primary. Material accreting onto a growing planetesimal should also produce ejecta with velocity dispersion on the order of the planetesimal's escape velocity $V_{escape}$. If, as the low 9 ± 2 Myr age of the system (Stauffer *et al.* 1995, Cameron *et al.* 2015) and the ring's planetary-scale minimum mass suggest, the ring is associated with a recent bout of planet formation within the last few Myr in HR4796A, then the mass and hence the escape velocity of the forming planet must be very low, << $M_{Mars}$, in order to keep the ring narrow, in contradiction to the sensible mass we find in the small dust particles of the belt. A possible solution to these apparent dynamical difficulties is that multiple planetesimal cores are growing in the HR 4796A cold comet dust ring (Rodigas *et al.* 2015, Perrin *et al.* 2015), as expected for forming outer giant planets (Izidoro *et al.* 2015) and these cores are serving to sheperd the ring material into the narrowly confined structure seen (but perhaps allowing ~$10^{-8}$ of the ring's particles through to spiral into, via P-R drag, the ~1 AU inner sublimation zone).

An alternate possibility, that the outer cold dust ring is the sublimating edge of the system's primordial evaporating proto-planetary disk, seems to be ruled out by the ring's narrowness in the radial dimension and the non-detection of any CO or PAH gas emission lines in our NIRDS HR 4796A spectrum or UV/mm studies of the system (Chen & Kamp 2004, Greaves *et al.* 2000, 2016), and our finding that the ring material is sourced from processed, devolatilized cometary material. It is also apparently ruled out by the similarly narrow rings found in the much older (~few 100 Myrs) Fomalhaut and HD 32297 systems (White *et al.* 2017, Asensio-Torres *et al.* 2016, and references therein), which long ago lost their primordial disks. Our NIRDS spectra show that Fomalhaut and HD 32297 do not harbor redly reflecting dust - likely because their rings are much farther out, and thus much colder, so they still harbor stable icy comet nuclei, consistent with the finding of CO gas associated with the dust rings of both these systems (Greaves *et al.* 2016, Matra *et al.* 2017). An important corollary of this statement is that we expect the planetisimals growing in the HR 4796A outer ring to be rocky, as their feedstock is devolatilized, as argued by Chen & Kamp (2004), while the planetisimals in the Fomalhaut and HD 32297 rings have abundant icy material in their feedstocks and so should be growing KBO and ice giant offspring.

# 5. Conclusions.

Using the NASA/IRTF SpeX 0.8-5.0 μm spectrometer we have obtained new R~2000 NIR observations of the HR 4796A debris ring system. Compared to other disk systems in our NIRDS survey of 50+ objects, we find a uniquely red excess flux increasing slowly with increasing wavelength from 0.8-5.0 μm. Combining the SpeX measurements with archival Spitzer IRS spectroscopy we find the slow red excess trend extending out to ~9 μm, where thermal emission from cold, ~100K dust in the system's outer ring at ~75 AU takes over. Normalizing the excess absolute level to the photometry of Debes *et al.* (2013), Rodriguez *et al.* (2015), and Milli *et al.* (2017), we find the excess consists of a strong NIR reflectance spectrum from the fine dust ring which is as red as the reddest Solar System objects, matching that of old, active comet nuclear material in our Solar System, plus a tenuous thermal emission component from close-in, T ~ 850 K circumstellar material evincing a organics + silicate emission feature complex at 7 - 13 um. No associated 20 μm silicate features are seen, implying that they must be sourced by the tenuous hot component. Unusual, weak emission-like features due to Si I, S I, Sr II, and Ca II were found in in the SpeX data, likely sourced by rocky dust evaporating in the 850 K hot component. An empirical cometary dust phase function (Schleicher & Marcus 2010) at i = 17$^o$ can reproduce the scattered light excess and 1:5 balance of scattered vs. thermal energy for the ring for an optical depth $\langle\tau\rangle \geq 0.10$ in a narrow belt with ± 4 AU vertical height (Perrin *et al.* 2015) and associated minimum dust mass of 0.1-0.7 $M_{Mars}$. Our results are consistent with an HR 4796A system consisting of a narrow sheparded belt of devolatilized cometary material associated with multiple rocky subcores forming a planet, and a small steady stream of dust inflowing from this belt to a rock sublimation zone at ~1 AU from the primary. These subcores were built from comets that have been actively emitting large, reddish dust for > 0.4 Myr at ~100K (the temperature at which cometary activity onset is seen in our Solar System).

## 6. Acknowledgements

The SPEX data used in this work were obtained by the authors as Visiting Astronomers at the Infrared Telescope Facility, which is operated by the University of Hawaii under Contract with the National Aeronautics and Space Administration, Science Mission Directorate, Planetary Astronomy Program. Our NIRDS observations take advantage of and add to the SpeX spectral library of ~200 cool FGKM stars (Rayner *et al.* 2009), and we are deeply indebted to J. Rayner for building SpeX and for providing his observing time and expertise to this project. The authors would like to thank T. Rodigas and M. Perrin for their many inputs and suggestions that have made the paper more robust. The authors wish to recognize and acknowledge the very significant role and reverence that the summit of Mauna Kea has within the indigenous Hawaiian community; we are most fortunate to have the opportunity to conduct observations from this mountain. C. Lisse would also like to gratefully acknowledge the support and input of the NASA Nexus for Exoplanet System Science (NExSS) research coordination network sponsored by NASA's Science Mission Directorate. R. Vervack and Y. Fernandez gratefully acknowledge their support, in part, by NSF grant AST-1109855 in performing this work.

## 8. Figures

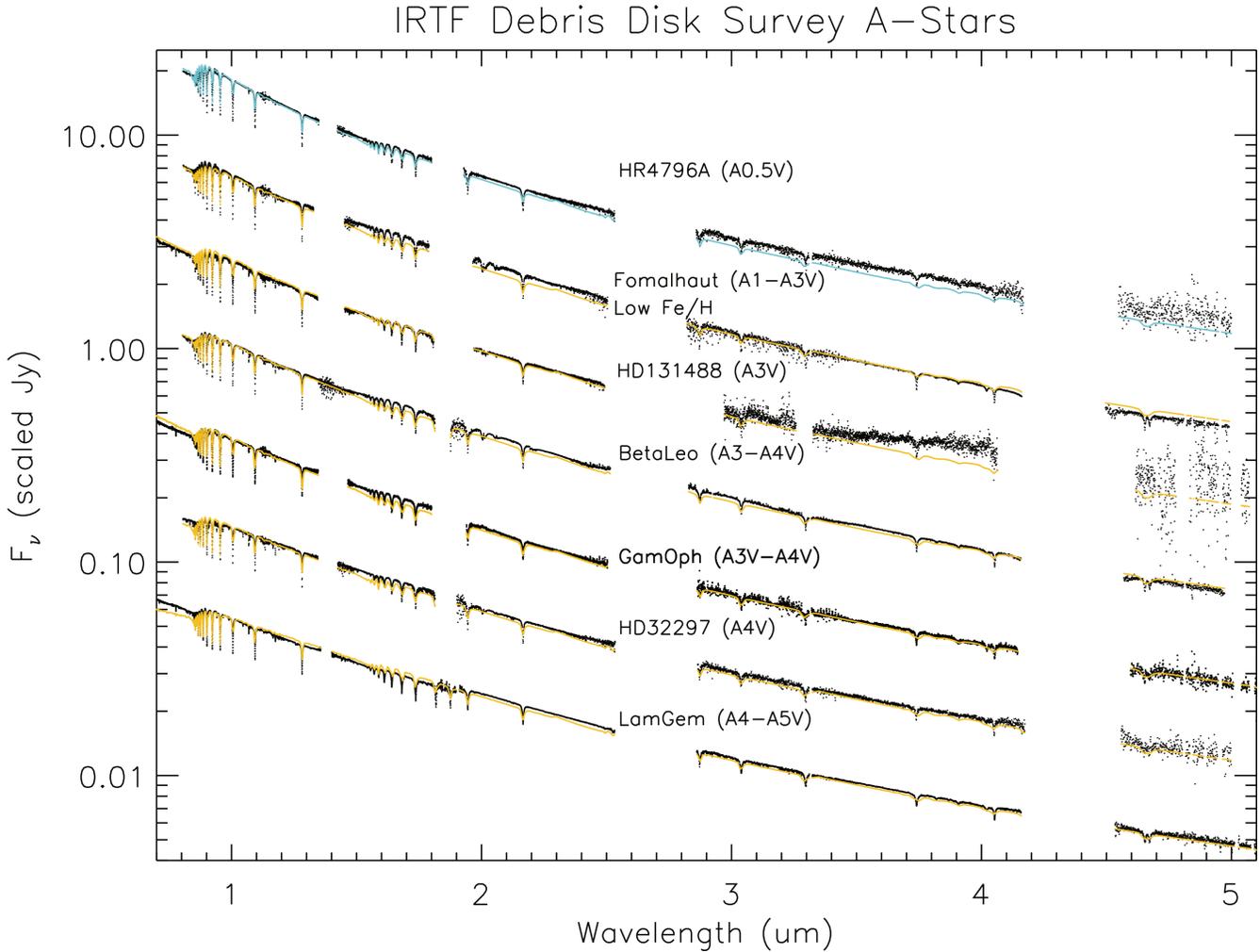

Figure 1 - *(a)* Early main sequence A-stars with reported circumstellar material observed from the NASA IRTF under the NIRDS program. R~1000, 0.8-5.0 μm SpeX data are in black and photospheric spectral models are in red. The typical spectral behavior for our program systems matches the photospheric model well from 0.8 - 1.3 um, exceeds it slightly (if at all) from 1.3 - 3 um, then matches it again from 3 - 5 um. HD 131488 shows the typical growing exponential behavior seen for other NIRDS stars (e.g., HD 113766, HD 15407A, HD 23514; Lisse *et al.* 2015, 2018) dominated by warm (~300K) thermal dust emission observed on the Wien law side of its SED. The spectral behavior of HR 4796A in the near-infrared, with its linearly growing excess as wavelength increases, is unique.

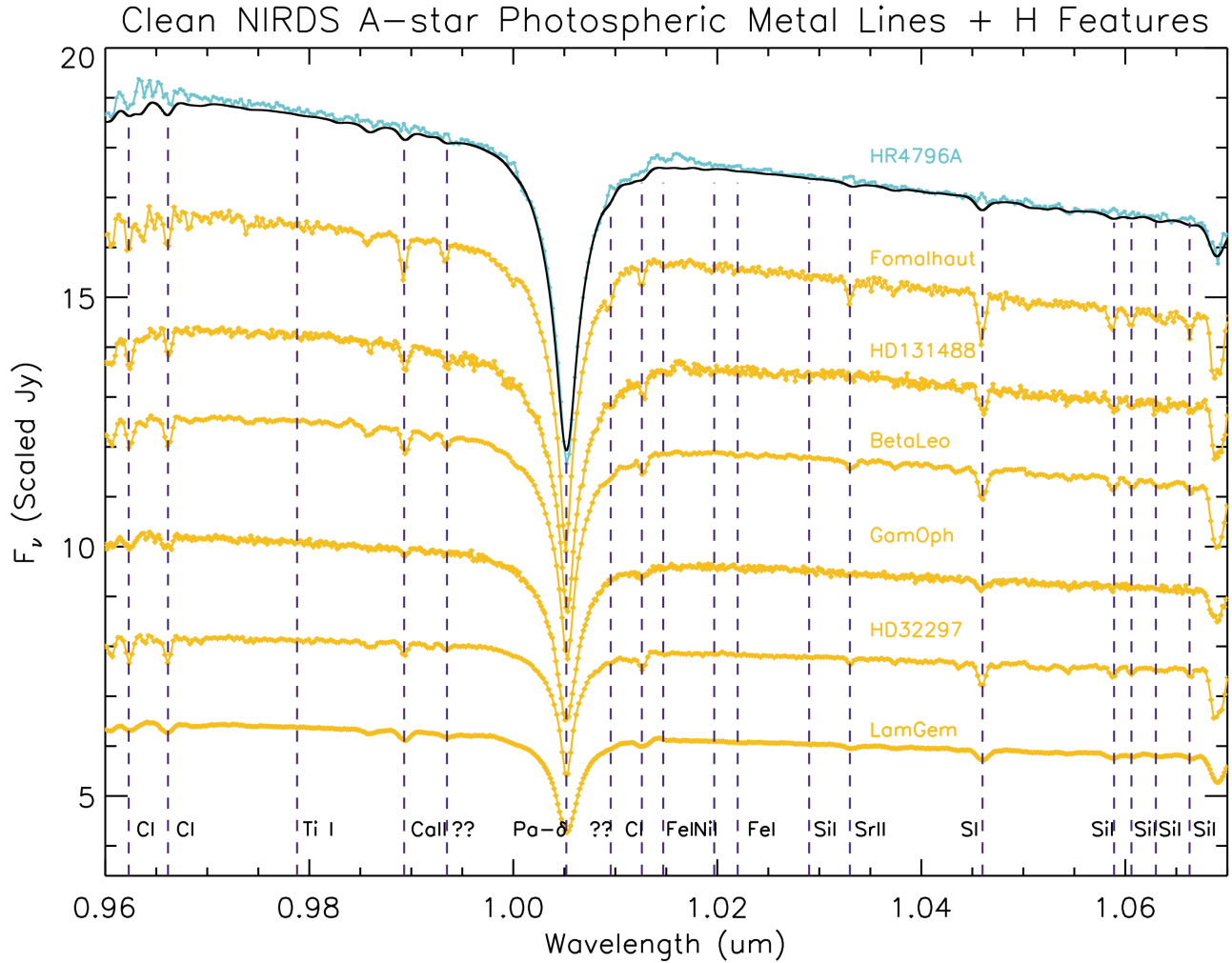

Figure 1 - **(b)** HR4796A SpeX spectrum in the 0.96 - 1.07 μm region (cyan) compared to those obtained for the other NIRDS A-stars of Figure 1(a) (gold), and versus our best-fit A0.5V stellar line model (black). Structure in these spectra are not due to noise, but to unresolved/unidentified lines. Small but significant emission-like features are seen for HR4796A in the Sr II and SI lines at 1.033 and 1.046 μm, respectively, where absorption features should lie. The Ca II line at 0.9893 μm and the Si I absorption lines at 1.0589 and 1.0606 μm also seem muted for HR4796A, most likely due to additional emission filling in the absorption features, but this is near the limit of our sensitivity to detect. By contrast, the Ti I and CI lines of HR4796A appear normal versus the other A-stars and the model. Unidentified emission features at 0.9935 and 1.00955 μm also seem to be present in the HR4796A spectrum.

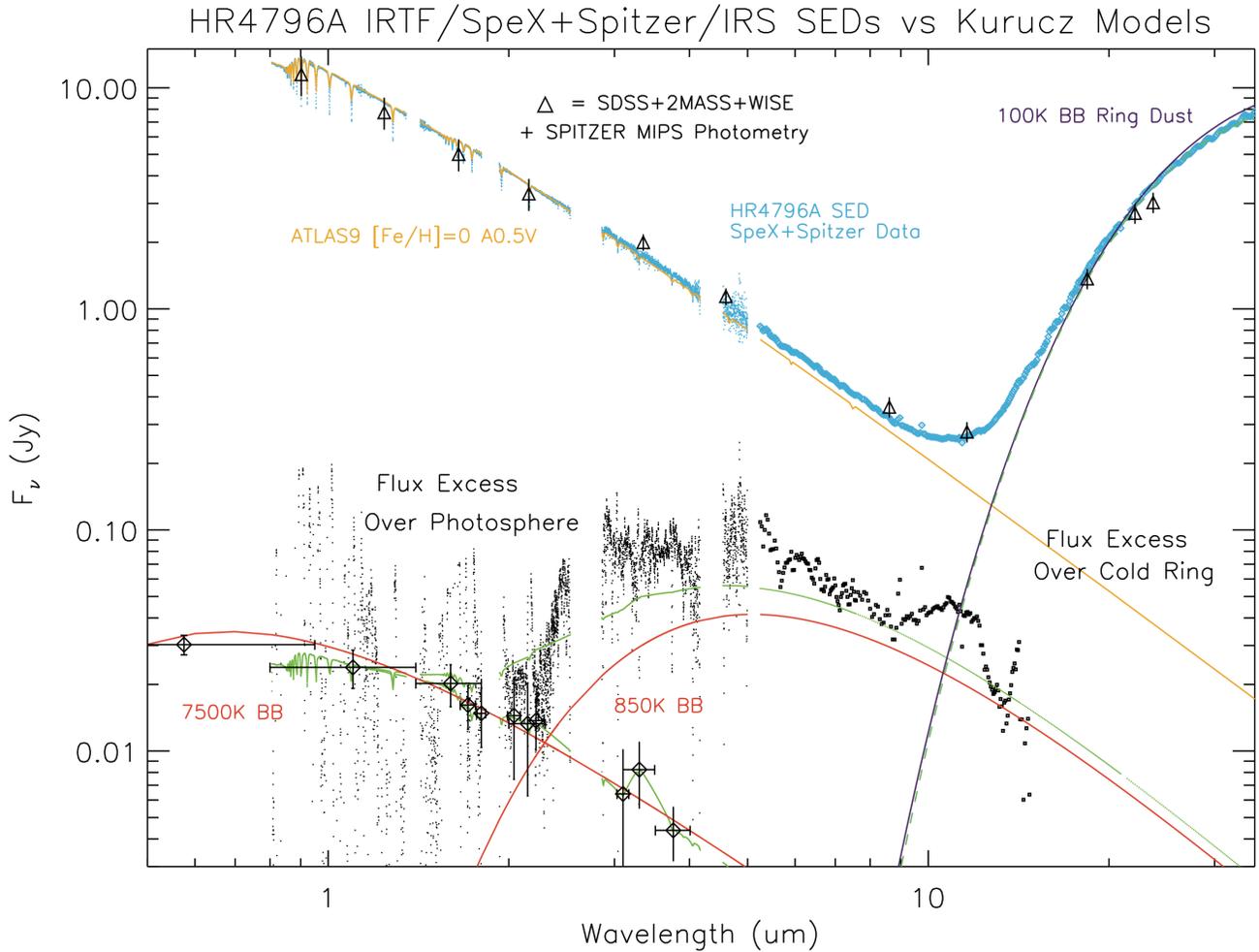

**Figure 2** - *(a)* Combined SpeX 0.8 - 5.0 and Spitzer 5.3 - 35 μm SED for HR 4796A (blue) compared to Vizier *Tycho, 2MASS, WISE, AKARI,* and *Spitzer* photometry (triangles with 2σ error bars). In yellow is the best-fit T = 9450K, g = 3.92 solar abundance ATLAS9 model matching our SpeX spectroscopy. The excess over the A0.5V model photosphere is plotted in black dots vs. the Rodigas *et al.* (2015) ring photometry (black diamonds). The shape of this excess at short wavelengths can be fit by thermal emission models with T ~ 7500K, at temperatures for which dust is not stable in solid form. The purple curve is a simple 100K blackbody fit to the outer dust ring's thermal emission. Removing the 100K model from the Spitzer produces the black boxes in the lower right and the signature of an 8 - 12 μm broad emission peak. **In the MIR where the spectra meet**, the red curve shows the flux from a new, ~850 K thermal component. The green curve shows the scattered ring flux plus the maximal allowed 850 K blackbody thermal emission.

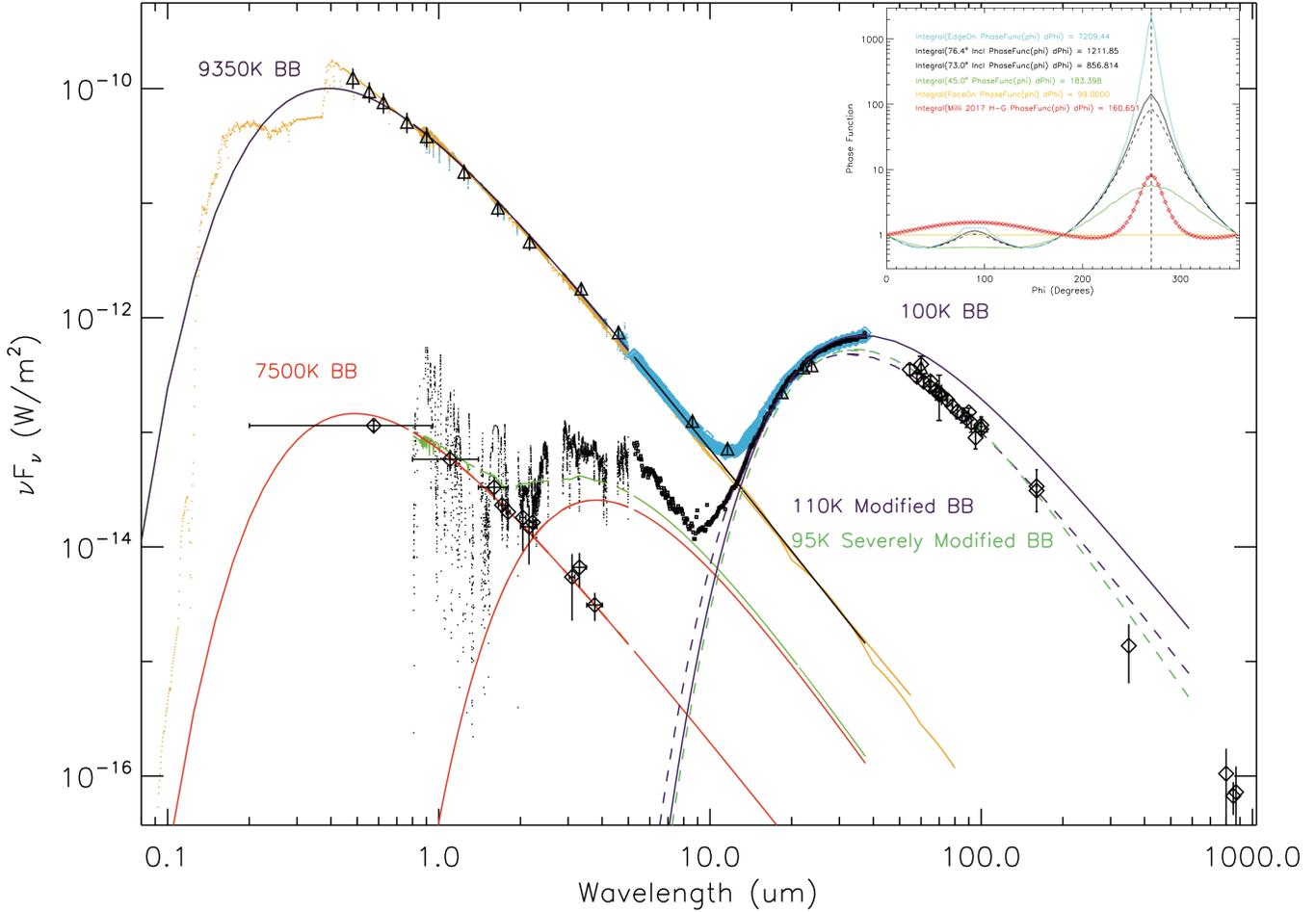

*(b)* - Broader 0.8 - 100 μm total SED, and the same two excess flux residuals shown in Fig 2a. Comparison of the NIR excesses to the FIR excess shows that there is 4 times as much energy in the FIR excess. The best FIR one-blackbody fit has T = 100 K as the dominant emission source. This thermal emission model predicts much more flux than is found by Spitzer MIPS, IRAS, and AKARI at 70 - 100 μm. The equilibrium temperature for a blackbody at 75 AU from a 23 $L_{Sun}$ A0.5V primary star is 71 K, suggesting the ring material has radius a < 50 *μm* and is superheating to a color temperature of ~100K in the MIR. The lack of strong 10 μm silicate emission features implies dust with a > 10 *μm,* above the blowout limit of 5 μm (Jura 1993). We must be observing dust and not a solid "wall" of comet nuclei, as only dust has the strongly peaked forward scattering behavior (Schleicher 2010; ***inset***) to redirect ~1.2 % of the primary's light with an 1:5 excess scattered light/thermal emission energy ratio.

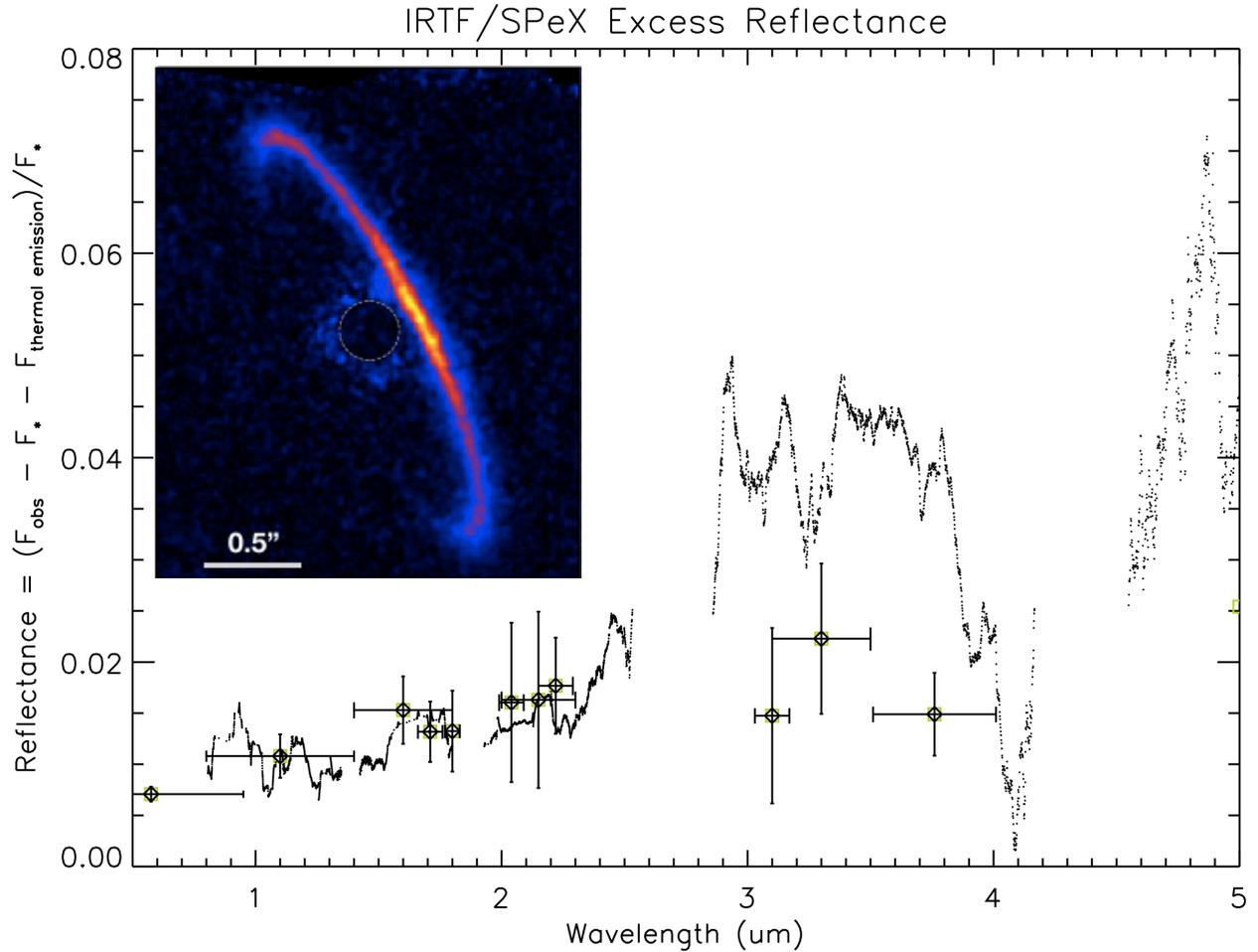

**Figure 3** - Spectral dependence of the SpeX excess reflectance for HR 4796A. Plotted in black dots are the residuals ($F_{observed}$ - $F_{ATLAS9\ model}$ - 850K BB)/$F_{ATLAS9\ model}$. Assuming the observed flux is $F_{observed}(\lambda) = F_{photosphere}(\lambda) + F_{scattered}(\lambda) = F_{photosphere}(\lambda) + \text{Reflectance}(\lambda) * F_{photosphere}(\lambda)$, the difference between the observed and photospheric modeled fluxes, normalized by the photospheric flux is the excess reflectance. The reflectance has been smoothed using a 100-point moving average. The excess reflectance is also consistent with the spectrophotometric run with wavelength of the ring's surface brightness at 90° phase found by Rodigas *et al.* (2015; black diamonds). Potentially interesting low-frequency features in the spectrum at ~3.4 μm due to the cometary C-H organic stretch (Quirico *et al.* 2016) and another predicted by Debes *et al.* at 3.8 - 4.0 μm for tholin material are also possibly present, but lie in a spectral region that is complicated by multiple strong telluric atmosphere emission lines (See Section 4). High-frequency narrow features are due to incomplete removal of photospheric emission lines and should be ignored. ***Inset:*** GPI 2 μm polarized light image of HR 4796A showing the strong forward scattering behavior of the ring (after Perrin *et al.* 2015).

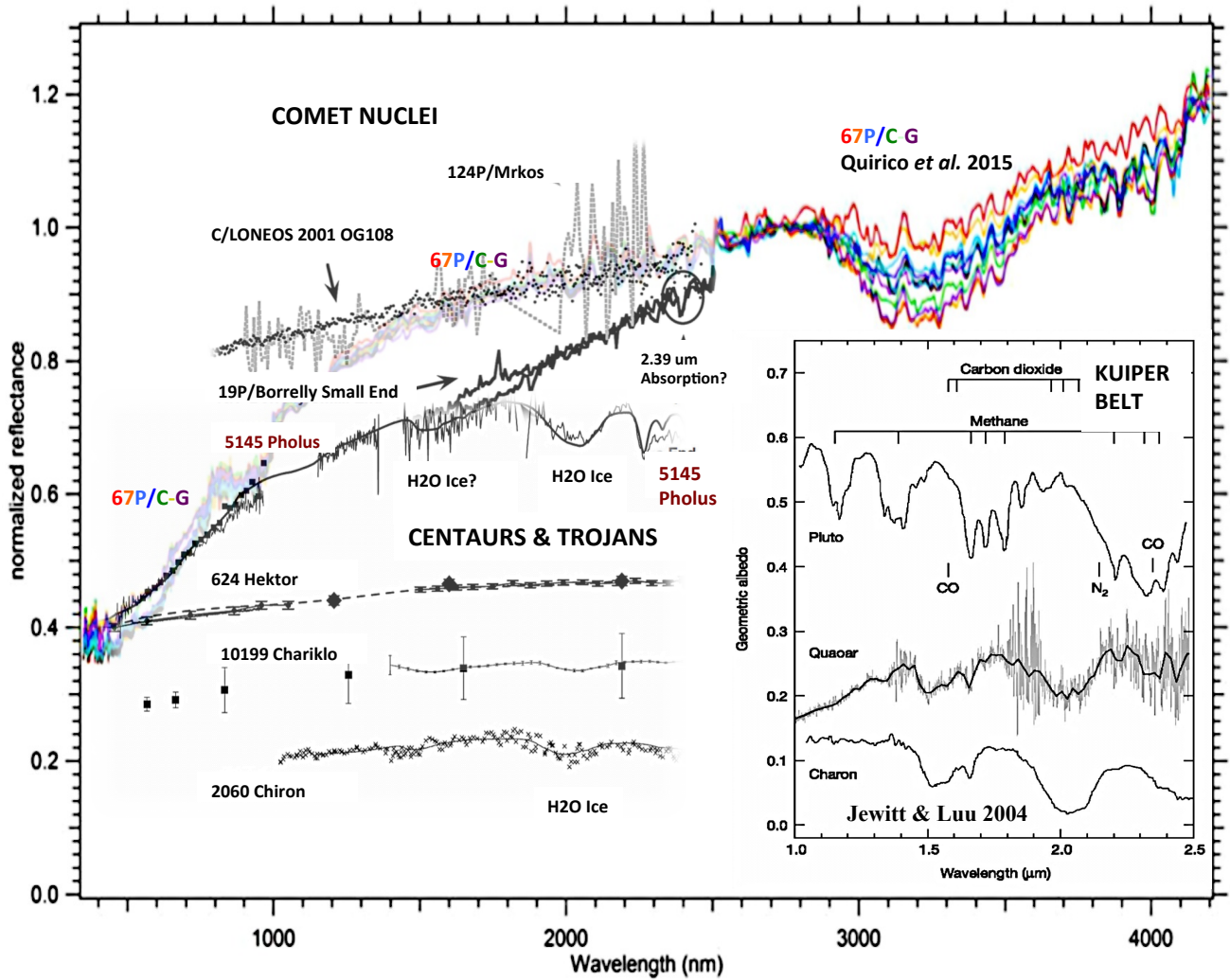

**Figure 4 -** NIR reflectance spectra for Solar System comets, Centaurs, and Trojan small bodies, the reddest objects in the NIR. Spectra are from Abell *et al.* 2005, Barucci *et al.* 2003, Licandro *et al.* 2003, Soderblom *et al.* 2004, Sunshine *et al.* 2006, and Quirico *et al.* 2015. Note that as a body becomes more active, it tends to become redder in the NIR, likely due to the loss of overbearing bluish ices and exposure of dark red organic-rich material beneath. Over the wavelength range 0.8 - 4.2 μm shown here, our HR 4796A SpeX reflectance spectrum (Fig. 2) is only consistent with the reflectance trends found for old, processed comet nucleus surfaces. This includes the possible absorption features at 2.4 and 3.4 μm (broad). Centaur Pholus, commonly known as the 'reddest object in the solar system', is only very red from 0.4 - 1.5 um; longwards of 1.5 μm $H_2O$ ice absorption on its surface takes over and the reflectance becomes flat with increasing wavelength. (***Inset***) Typical KBO NIR reflectance spectra, dominated by surface ice absorptions, are grey to slightly blue in the NIR. (After Jewitt & Luu 2004, Grundy *et al.* 2016).

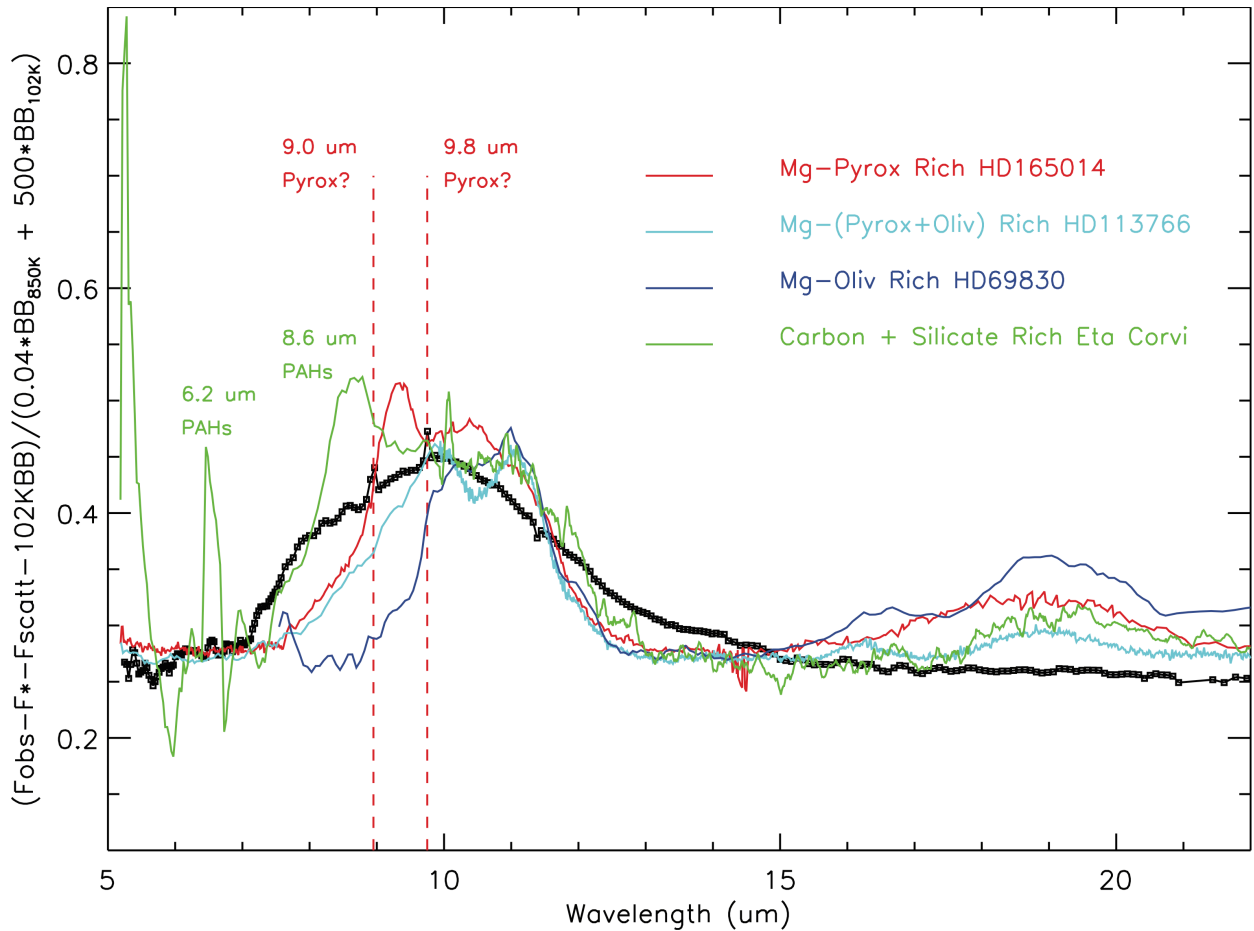

**Figure 5 - Carbon + silicate-rich emission feature created by taking the Spitzer/IRS 5-20 μm residual of HR4796A dividing it by the predicted dust thermal blackbody behavior (black).** The feature is compared to the silicate features found for the extremely pyroxene rich debris disk of HD165015 of Fujiwara *et al.* 2010 (red), the extremely olivine rich HD69830 disk of Beichmann *et al.* 2005, 2011 (purple), the feature produced by the ~ 1:1 pyroxene: olivine dust of HD113766 (aqua; Lisse *et al.* 2008); and the feature produced by the carbon- and pyroxene-rich KBO dust of the η Corvi debris disk (light green; Lisse *et al.* 2012). The dashed lines are the wavelength locations of disk and comet pyroxenes, as determined by meteoritic and laboratory analogues (Morlok *et al.* 2014). The narrow HR4796A peaks at 9.0 and 9.8 um, while expected for pyroxenic material, are not due to single point defects but are still puzzlingly strong. Neither the pyroxene-dominated HD165014 nor the olivine-dominated HD69830 dust spectra are good matches for the HR4796A silicate feature. Some kind of dust like the η Corvi population (but without any gaseous PAHs) seems to be the best match, consistent with the source of the hot dust arising in primitive outer solar system bodies like those resident in the HR4796A cold dust ring.